%% file: main.tex
\documentclass[12pt]{article}
\usepackage[letterpaper,margin=1in,headsep=0.1in,headheight=1in]{geometry}

\usepackage[T1]{fontenc}
\usepackage{ae,aecompl}
\usepackage[utf8]{inputenc}
\usepackage{lipsum}

\usepackage[hidelinks]{hyperref}
\usepackage{graphicx}
\usepackage[sort&compress,numbers]{natbib}

\usepackage{siunitx}
\usepackage{amsmath,amssymb,amstext}
\usepackage{xspace}
\usepackage{aas_macros}

\usepackage{enumitem}
\usepackage{xcolor,colortbl,multirow}
\usepackage{graphicx}
\usepackage{comment}
\newcommand{\mc}[2]{\multicolumn{#1}{c}{#2}}
\newcommand{\mcl}[2]{\multicolumn{#1}{l}{#2}}

\definecolor{MyRed}{rgb}{1,.45,.35}
\definecolor{MyPurple}{rgb}{.55,.5,.7}
\definecolor{MyBlue}{rgb}{.45,.62,.81}
\definecolor{MyYellow}{rgb}{1,.85,.35}
\definecolor{MyGreen}{rgb}{.5,.7,.7}

\newcommand{\annika}[1]{{\textcolor{green}{\sf{[AHGP: #1]}}}}

\newcommand\snowmass{
\begin{center}
  \rule[-0.2in]{\hsize}{0.01in}\\
  \rule{\hsize}{0.01in}\\
  \vskip 0.1in
  Submitted to the Proceedings of the US Community Study\\ 
  on the Future of Particle Physics (Snowmass 2021)\\
  \rule{\hsize}{0.01in}\\
  \rule[+0.2in]{\hsize}{0.01in}\\[-2em]
\end{center}
}

\usepackage[firstpage=true]{background}
\backgroundsetup{contents={\parbox{6.5in}{\snowmass}}, scale=1,placement=top,opacity=1,color=black,position={3.25in,1.2in}}

\usepackage{fancyhdr}
\fancypagestyle{plain}{%
  \fancyhf{}%
  \fancyhead[C]{}
  \fancyfoot[C]{\thepage}
}

\fancypagestyle{empty}{%
  \fancyhf{}%
  \fancyhead[C]{{\it Cosmological Simulations for Dark Matter Physics}}
  \fancyfoot[C]{\thepage}
}
\title{Snowmass2021 Cosmic Frontier White Paper:\\
    Cosmological Simulations for Dark Matter Physics}
\date{}

\usepackage{authblk}

\author[1,2,$\dagger$]{Arka Banerjee}
\author[3]{Kimberly K.~Boddy}
\author[4]{Francis-Yan Cyr-Racine}
\author[5]{Adrienne L. Erickcek}
\author[6]{Daniel Gilman}
\author[7]{Vera Gluscevic}
\author[8]{Stacy Kim}
\author[9,10]{Benjamin~V.~Lehmann}
\author[11]{Yao-Yuan Mao}
\author[12]{Philip Mocz}
\author[13,$\dagger$]{Ferah Munshi}
\author[14,7]{Ethan~O.~Nadler}
\author[15]{Lina Necib}
\author[16]{Aditya Parikh}
\author[17,$\dagger$]{Annika H. G. Peter}
\author[18]{Laura Sales}
\author[15]{Mark Vogelsberger}
\author[19] {Anna C. Wright}
 
\affil[1]{Department of Physics, Indian Institute of Science Education and Research,
Homi Bhabha Road, Pashan, Pune 411008, India}
\affil[2]{Fermi National Accelerator Laboratory, Batavia, IL 60510, USA}
\affil[3]{Department of Physics, The University of Texas at Austin, Austin, TX 78712, USA}
\affil[4]{Department of Physics and Astronomy, University of New Mexico, Albuquerque, NM 87106, USA}
\affil[5]{Department of Physics and Astronomy, University of North Carolina at Chapel Hill, Phillips Hall CB 3255, Chapel Hill, NC 27599, USA}
\affil[6]{Department of Astronomy and Astrophysics, University of Toronto, Toronto, ON M5S 3H4, CA}
\affil[7]{Department of Physics \& Astronomy, University of Southern California, Los Angeles, CA, 90007, USA}
\affil[8]{Department of Physics, University of Surrey, Guildford GU2 7XH, UK}
\affil[9]{Department of Physics, University of California, Santa Cruz, Santa Cruz, CA 95064, USA}
\affil[10]{Santa Cruz Institute for Particle Physics, Santa Cruz, CA 95064, USA}
\affil[11]{Department of Physics and Astronomy, Rutgers, The State University of New Jersey, Piscataway, NJ 08854, USA}
\affil[12]{Lawrence Livermore National Laboratory, Livermore, CA 94550, USA}
\affil[13]{Department of Physics and Astronomy, University of Oklahoma, 440 W. Brooks St, Norman, OK 73019, USA}
\affil[14]{Carnegie Observatories, 813 Santa Barbara Street, Pasadena, CA 91101, USA}
\affil[15]{Department of Physics \& Kavli Institute for Astrophysics and Space Research, Massachusetts Institute of Technology, Cambridge, MA 02139, USA}
\affil[16]{Department of Physics, Harvard University, Cambridge, MA 02138, USA}
\affil[17]{CCAPP, Department of Physics, Department of Astronomy, The Ohio State University, 191 W. Woodruff, OH 43210, USA}
\affil[18]{Department of Physics and Astronomy, University of California, Riverside, CA 92521, USA}
\affil[19]{The William H. Miller III Department of Physics \& Astronomy, Johns Hopkins University, 3400 N. Charles St, The Bloomberg Center, Baltimore, MD 21218, USA}
\affil[$\dagger$]{Paper Facilitator}
\begin{document}

\maketitle

\begin{abstract}

Over the past several decades, unexpected astronomical discoveries have been fueling a new wave of particle model building and are inspiring the next generation of ever-more-sophisticated simulations to reveal the nature of Dark Matter (DM).  This coincides with the advent of new observing facilities coming online, including JWST, the Rubin Observatory, the Nancy Grace Roman Space Telescope, and CMB-S4. The time is {\it now} to build a novel simulation program to interpret observations so that we can identify novel signatures of DM microphysics across a large dynamic range of length scales and cosmic time.  This white paper identifies the key elements that are needed for such a simulation program. We identify areas of growth on both the particle theory side as well as the simulation algorithm and implementation side, so that we can robustly simulate the cosmic evolution of DM for well-motivated models. We recommend that simulations include a fully calibrated and well-tested treatment of baryonic physics, and that outputs should connect with observations in the space of observables. We identify the tools and methods currently available to make predictions and the path forward for building more of these tools.  A strong cosmic DM simulation program is key to translating cosmological observations to robust constraints on DM fundamental physics, and provides a connection to lab-based probes of DM physics.

%ADW: How about this as an Executive Summary before the Introduction?
\begin{comment}
\noindent The following priorities summarize this white paper:\\
\begin{enumerate}

    \item {\bf Close collaboration between simulators and particle theorists} to both identify key models and areas of parameter space and to successfully implement these models. This includes starting from appropriate initial conditions to providing simulation outputs in a manner that is meaningful to particle theorists and simulators alike.\\

    \item {\bf Algorithm development and code comparison tests} ensuring that simulations meet the required precision targets set by the sensitivity of the new facilities. For hydrodynamic modeling, this includes evaluation and comparison of different subgrid physics parameterizations. \\

    \item {\bf Performing simulations with full hydrodynamics with validated subgrid models and numerical resolution} at the relevant redshifts and cosmological scales.\\

    \item {\bf Analysis of simulation outputs in the realm of observations.}  Forward modeling to the space of observables to enable apples-to-apples comparison between model and data.\\

    \item {\bf Fast realizations of observables for inference of DM properties} in order to constrain DM particle parameters from observation on feasible timescales.\\

    \item {\bf Identifying novel signatures from simulations and guidance to observers} derived both from numerical simulations and fast realizations that point to signatures of DM physics. \\

%\lipsum[1]
\end{enumerate}
\end{comment}
\end{abstract}

% Body of the text
\section{Introduction and Executive Summary}

\input{intro_new}

\section{Opportunities}
%\ADW{Snowmass is not about identifying "Priorities". How about "Opportunities"?}
A successful program to extract DM physics from cosmological observations depends critically on accurate theoretical predictions.  Here, we propose a pathway toward achieving the goal of characterizing DM physics with cosmological observations through the robust mediator of a simulation program.  

\subsection{Need \#1: Collaboration between cosmological simulators and particle theorists}\label{sec:need1}

\noindent \textbf{Current Status:} By its very nature, the formation of structure in our Universe involves physical processes spanning a large dynamical range. The interactions between particles on the microphysical scale can directly impact the largest astrophysical and cosmological scales observable. This enormous difference in scales between the models particle theorists write down and the observations made by astronomers means that some form of coarse-graining (see next section) is absolutely necessary to map out how structure forms within any possible DM scenario.
%To avoid pitfalls and have realistic expectations, it is important for theorist to be aware of how such coarse-graining is done.
%\annika{Summarizing the key point: physics on the smallest scales affects the largest scale structure we can measure.  Need care in mapping the physics of small scales to the phenomena on large scales.  Need a ``mediator" framework to bridge the two.}

The huge dichotomy in scales between particle physics and astrophysics also means that structure formation is not necessarily sensitive to specific microscopic Lagrangian parameters such as masses or couplings, but rather to key combinations of parameters determining macroscopic quantities such as interaction rates, cutoff scales, or halo core sizes. The work of theorists is important for identifying these important parameter combinations within any DM theory, as these are the quantities that can ultimately be constrained by observations. Crucially, this mapping from DM microphysics to quantities relevant to astrophysical structure formation allows the classification of DM models according to their structure formation properties, enabling a limited number of simulations to represent models that might have very different particle origins. For instance, the absence of a measurable suppression of structure on small scales could be interpreted as a lower bound on the warm DM thermal mass, or as an upper bound on the cross section between DM and a dark thermal bath.  This is powerful as it allows simulators to cover a broad range of DM model space with a computationally realistic number of simulations. %\annika{Key point summary: need to map microphysics to phenomenological parameters that affect structure formation.  For any specific phenomenological parameter, many microphysical parameters map to it.  Helpful for reducing simulation demands.}

DM microphysics affects structure formation in a multitude of ways, and the combination of the different physical effects could lead to unique observational signatures. For example, a given DM model might predict both a specific shape for the initial matter transfer function and some important physical processes in the nonlinear regime such as self-interaction or dissipation. The combination of these effects can lead to important impacts on observables that might not be present if either effects were considered separately, hence providing new insights on the DM microphysics. Close collaborations between theorists and simulators is necessary to identify models and area of parameter space where such interplay between different physical effects might be present and explore them in simulations. As an example of such collaboration, Ref.~\cite{vogelsberger2016} self-consistently explored through simulations a category of models in which both dark acoustic oscillations (affecting the initial transfer function) and self-interaction are present, unveiling an important interplay between these two effects on the density profiles of halos. In parallel, Ref.~\cite{2016PhRvD..93l3527C} classifies such theories in terms of their structure-formation properties, allowing one to map the simulation results to specific corners of particle parameter space. This effective mapping, known as the ``Effective Theory of Structure Formation'' (ETHOS) framework, has been successfully used to systematically explore the role of DM physics on a range of observables \cite{Lovell:2017eec,Bose:2018juc,Bohr:2020yoe,Munoz:2020mue,Bohr:2021bdm}.

%\annika{good.  SHould give more play to FYCR's ETHOS framework as an example!  Present as a success.  Also be clear what the challenges are, as well as the opportunities if we get through the challenges.} 

%\bvl{implementation ``section''}

\noindent \textbf{Future Opportunities:} Beyond selection of representative models, continued collaboration between theorists and simulators is key to the successful \textit{implementation} of particular simulations. This is needed even at the stage of generating appropriate initial conditions for the simulations. Indeed, in many models, unique predictions arise largely from modifications to the transfer function. For instance, warm DM models generically produce a small-scale cutoff in the power spectrum, which has been well studied in many simulations, but such scenarios also exhibit nontrivial model-dependent phase space distributions, which are more difficult to explore generically. Going beyond thermal distributions calls for detailed examination of initial conditions in concrete models, as in Ref.~\cite{Venumadhav:2015pla}. Understanding these features and their relative importance must be a shared responsibility of the theory and simulation communities.
%\annika{can cite Francis-Yan's paper w/Teja, Kev, Chris.  Good example.  Another is SIDM and what type of cross section matters for structure formation.} 

After fixing initial conditions, theorists still have a significant role in negotiating the compromises necessary for practical implementation. First, theorists must identify the most flexible formulation of the physical processes of interest, so that a limited suite of simulations can be applied to the broadest possible set of models. Second, while simulation experts understand the constraints of available software and computational resources, theorists are well-positioned to propose or evaluate approximation schemes for the physical processes of interest: the input of theorists is needed to determine which effects can be neglected and which cannot. Thus, simulations of novel classes of DM physics will have the best foundation for success if theorists are directly involved in the algorithm design choices and testing. %\annika{Some overlap w/need 2 -- need to figure out where to put what ideas.  I think the point about making sim products accessible to particle physicists belongs here, though.}

A key ongoing challenge well-suited to such collaboration is the implementation of simulations with SIDM. On the one hand, the space of SIDM models gives rise to self-interaction cross sections with a range of preferred scales and velocity dependences (such as interactions via light mediators~\cite{Tulin:2013teo}, atomic-like DM~\cite{CyrRacine:2012tfp,Boddy:2016bbu}, nuclear-like DM~\cite{Boddy:2014yra}), and to additional dynamical features associated with e.g.\ excited states or radiative transfer in the dark sector~\cite{ArkaniHamed:2008hhe,Schutz:2014nka,Blennow:2016gde,Das:2017fyl}. Different forms of interactions in the dark sector can lead to various velocity dependences in the effective cross-section. In addition, these velocity dependences can be modified significantly due to non-perturbative Sommerfeld enhancement depending on combinations of parameters such as coupling strength and ratios of mediator to dark matter mass. Furthermore, even the presence of this enhancement can depend on the nature of the underlying interactions~\cite{Agrawal:2020lea}. SIDM models give rise to a rich array of cross section behavior, depending closely on dark sector microphysics. On the other hand, implementation of SIDM in simulations is notoriously complex, and requires a choice of approximation scheme to map the effects of microscopic self-interactions to the mesoscopic scales of simulated particles.

%\bvl{utilization ``section''}

Not only is it important for particle theorists to guide simulators in incorporating relevant input from physically sound theories, but the results of simulations should be made available in a manner that is useful to theorists.
Clear communication about the data format of outputs, the units and/or normalizations used, etc.\ help make simulations accessible to the broader community.
It is also beneficial for studies to provide accessible information on how particle physics processes are coarse gained within simulations.
This communication between simulators and theorists is necessary for proper interpretation of simulation results and the scope of their impact on particle physics properties of DM. \\
%on the assumptions made and limitations in interpreting the results and their impact on particle physics.

\noindent \textbf{Summary:} Given the importance for simulators and theorists to interact efficiently to produce the most impactful science, it is crucial to support the work that strives to bridge gaps in expertise between (and within) the simulation and theory subfields.
Equally important is the support of work that connects simulation-based studies to astrophysical and cosmological observations, as well as complementary searches of DM, such as direct detection, indirect detection, and collider experiments.
As understanding the fundamental nature of DM continues to be one of the most prominent and outstanding questions in physics, progress will rely on the continued cooperation between theory, simulation, and experiment, each guiding the other. 
%\annika{and connect to other types of experiments--would be a good place to mention indirect detection or lab experiments, and point to need 6.  In any case, we need to zoom out and show how this connects to the broad landscape of DM detection.}

\subsection{Need \#2: Algorithm development and code comparison tests}\label{sec:need2}

\input{need2.tex}

\subsection{Need \#3: Simulations with full hydrodynamics for observational targets}\label{sec:need3}
%Writers: Stacy Kim, Mark Vogelsberger\\

\input{need3.tex}

\subsection{Need \#4: Comparison of observations and simulations should be carried out in the space of observables}\label{sec:need4}
%Writers:  Anna Wright, Laura Sales
%\item Comparisons between theoretical predictions and observational findings are complicated by the fact that observers and simulators use different pipelines to analyze their data. As a result, two methods that appear to be measuring the same parameter may actually be measuring two very different things, resulting in the appearance of a disagreement between predictions and observations that can be resolved via apples-to-apples comparisons. Use Fig 4 from Brooks et al., 2017 as an example

\noindent \textbf{Current Status:} A robust comparison between theoretical models and observations is often complicated by the fact that the properties that models best predict are rarely  direct observables. During the last decades we have learned that the correct translation between theoretical predictions and what they mean for the observational data is of paramount importance to determine the validity, or not, of our cosmological models. For instance, several of the strongest ``problems'' for $\Lambda$CDM can be resolved when a proper apples-to-apples comparison is established. 

%%%%%%%%%%%%%%%%%%%%%%%%%%%%%%%%%%%%%%%%%%%%%%%%%%%%%%%%%%%%%%%%%%%%%%%%%
\begin{figure}[t]
    \centering
    \includegraphics[width=.94\textwidth]{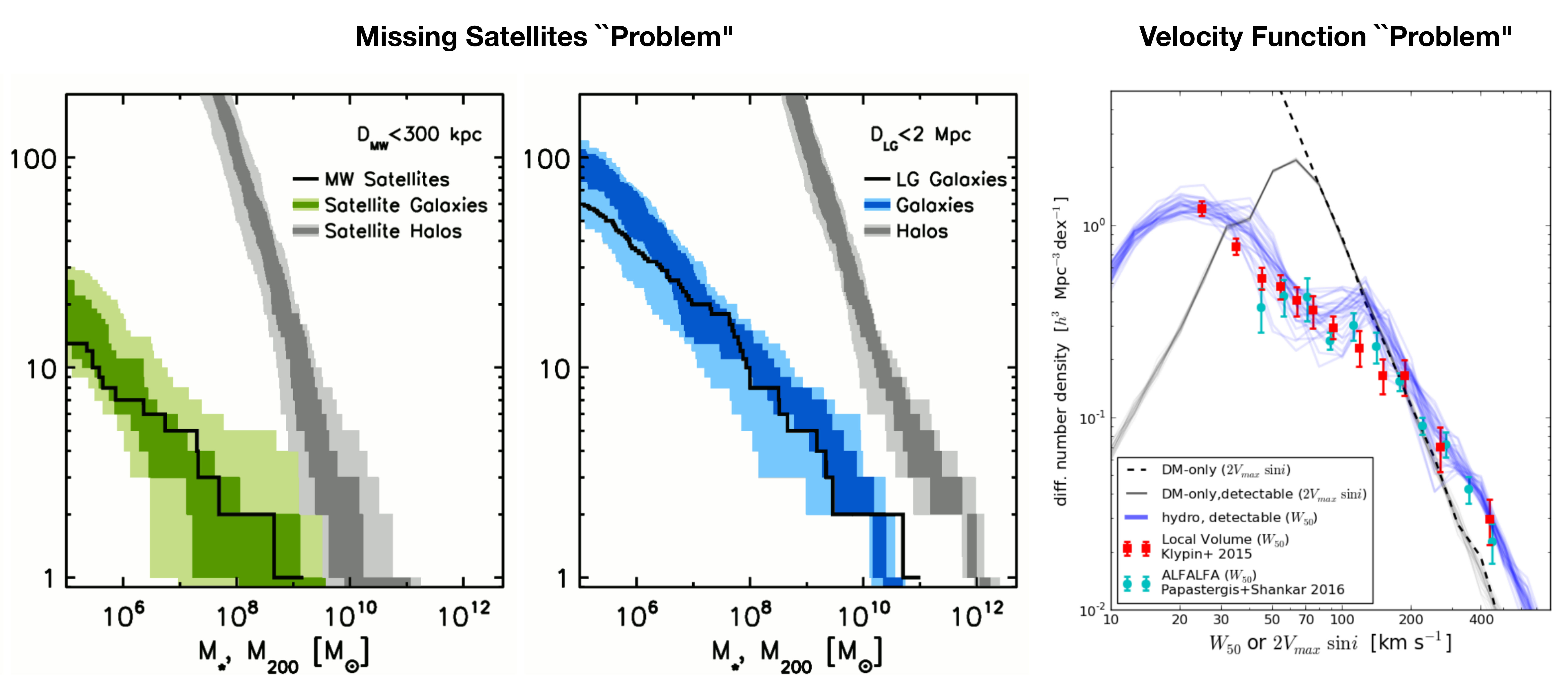}
    \caption{Examples where a proper comparison between simulations and observations has helped resolve purported ``problems" in CDM. {\it Left and middle panels}: modeling galaxy formation physics on top of N-body simulations helps resolve the missing satellites problem. Galaxy formation lowers the predicted number of subhalos in CDM (shown in gray) to the number of predicted luminous dwarf galaxies (green/blue), which are in good agreement with observations of the MW and the Local Volume  \citep{Sawala2016}. {\it Right}: in addition to galaxy formation physics, the translation from circular velocity function (solid gray and dashed lines) to HI-width function (blue lines) helps explain the shallow slope in the observed HI-width velocity function \citep{Brooks2017}.}
    \label{fig:sims_obs_comparison}
\end{figure}
%%%%%%%%%%%%%%%%%%%%%%%%%%%%%%%%%%%%%%%%%%%%%%%%%%%%%%%%%%%%%%%%%%%%%%%%%

%\annika{This is a good paragraph.  I especially like the HI part--doing ``observations" of simulations is extremely important!}
Such is the case of, for example, the ``missing satellites problem'' which early on pointed to a far lower number of dwarf satellites observed in the Milky Way compared to the number of DM subhalos predicted by N-body only simulations of Milky Way-like halos \citep[e.g.,][]{1999ApJ...524L..19M,1999ApJ...522...82K}. Several teams have now successfully demonstrated that this problem arises primarily because it compares two very different things: galaxies versus DM subhalos \citep[e.g.,][]{2013ApJ...765...22B}. Once the modeling of galaxy formation is coupled to that of the DM in order to predict the stellar content of those subhalos, predictions from CDM and the observed number of dwarfs can be reconciled (see Fig.~\ref{fig:sims_obs_comparison}). Similarly, the shallow low-velocity end of the velocity distribution, as measured by H\textsc{i}-width observations, has been considered to be in tension with the steep rise of the velocity function predicted by CDM \citep{2011ApJ...739...38P,2015MNRAS.454.1798K}. However, H\textsc{i}-widths are not a good proxy for circular velocity, the quantity traditionally predicted in theoretical models, in particular in the regime of dwarf galaxies. Works that not only take into account the baryonic physics but also make an effort to establish a fair comparison with observations \citep[e.g.,][]{2016MNRAS.463L..69M,Brooks2017} find good agreement between the observed HI-width distribution in observations and those predicted in CDM simulations (see Fig.~\ref{fig:sims_obs_comparison}). A careful comparison between predicted velocities in simulations and observations is also necessary for reconciling the Baryonic Tully-Fisher relation in dwarfs \citep[e.g.,][]{2016MNRAS.459..638B,2020MNRAS.498.3687G} and to understand the observed diversity of rotation curves \citep[e.g.,][]{2004ApJ...617.1059R,2015MNRAS.452.3650O,2016MNRAS.462.3628R,2019MNRAS.482..821O,2020MNRAS.495...58S}.

%\item To overcome this, simulators must make predictions in the space of observables. Summary of tools that can be used to convert simulation data to mock data for specific instruments (e.g., STIPS, powderday, SKIRT, ananke). Have a flow chart showing how simulation data can be translated into specific observables. \ferah{connect with need 5 examples}

\begin{figure}
    \centering
    \includegraphics[width=.94\textwidth]{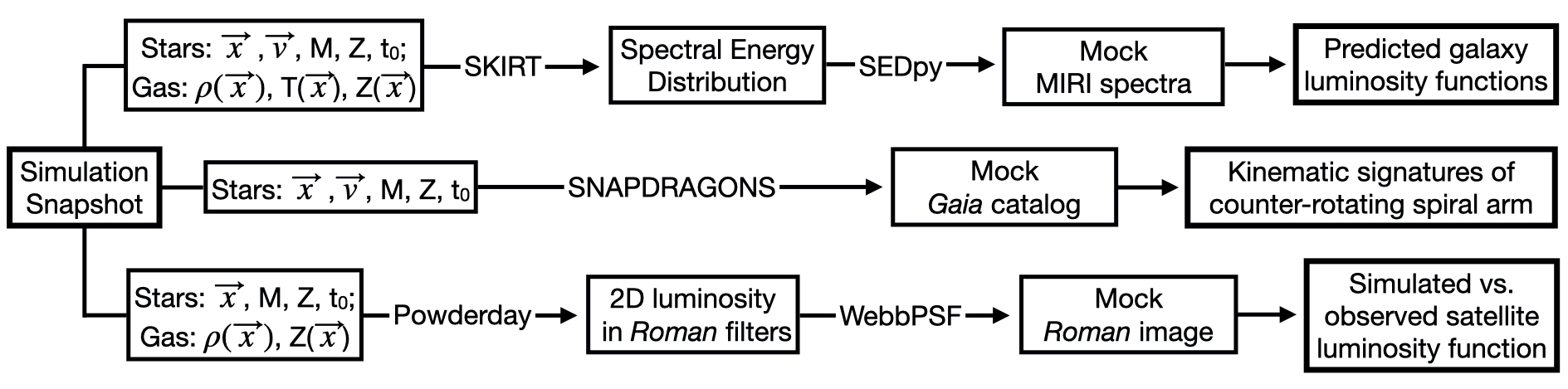}
    \caption{Examples of pipelines that can be used to convert outputs from numerical simulations into mock observations made with specific instruments. Although there are many post-processing suites (e.g., Powderday or SKIRT) available to add the effects of dust or other physics not included in simulations to data from simulation snapshots, outputs from these programs often need to be combined with instrument-specific software in order to include the effects of e.g., point spread functions. Examples of the top and middle pipelines can be found in \citep{2022MNRAS.tmp...24S} and \cite{2015MNRAS.450.2132H}, respectively.}
    \label{fig:mockobspipeline}
\end{figure}

\noindent \textbf{Future Opportunities:} The above examples highlight the power of proper theory-observation comparisons and the need to build solid tools to translate theoretical predictions to the land of observable quantities. Several such efforts have already been established and many have been made publicly available. Some of the most commonly used tools for galaxy formation studies are algorithms that take snapshots from hydrodynamic simulations as inputs and translate them into spectral energy distributions (SEDs) and images in user-specified filters. These include Sunrise \citep{2006MNRAS.372....2J,2010MNRAS.403...17J}, SKIRT \citep{2003MNRAS.343.1081B,2011ApJS..196...22B,2020A&C....3100381C}, and Powderday \citep{2021ApJS..252...12N}, all of which use Monte Carlo radiative transfer to incorporate the effects of e.g., dust absorption and scattering, which are often not included in situ in simulations. Outputs from these and similar software packages can also be combined with image simulators adapted to the specifics of a given telescope or mission (e.g., STIPS, WebbPSF \citep{2012SPIE.8442E..3DP,2014SPIE.9143E..3XP}, SEDpy \citep{2019ascl.soft05026J}) or packages to translate the 6D information of simulations into projected positions and kinematics (e.g., Galaxia \citep{2011ApJ...730....3S}, SNAPDRAGONS \citep{2015MNRAS.450.2132H}, Ananke \citep{2020ApJS..246....6S}) for the construction of mock surveys. These mock observation packages (see Figure \ref{fig:mockobspipeline} for several sample pipelines) can be used not only to enhance simulation data and convert it into a format that can be fed into observational data reduction pipelines, but also to incorporate typical observational uncertainties. This allows for a more accurate assessment of potential tension between theory and observations.

%%%%%%%%%%%%%%%%%%%%%%%%%%%%%%%%%%%%%%%%%%%%%%%%%%%%%%%%%%%%%%%%%%%%%%%%%
%%%%%%%%%%%%%%%%%%%%%%%%%%%%%%%%%%%%%%%%%%%%%%%%%%%%%%%%%%%%%%%%%%%%%%%%%

%\item Detail a few of the observables that hold most promise to constrain cosmology: internal dwarf galaxy kinematics (rotation curves, stellar/gas velocity dispersion); satellite and total luminosity functions; radial distribution of satellites; morphology and SFR, lensing, gaps in streams, Lyman-alpha forest, CGM?? (coordinate with Sec. 2.3 and previous item).

%\item Upcoming datasets that we should be preparing for: LSST, DES, Roman, Webb
While the physics of galaxy formation is extremely important in shaping the population of galaxies, the underlying DM physics that is assumed in a given cosmological model may leave clear imprints in the number, inner structure and the distribution of those galaxies. Currently, there are several observables that have been identified to hold the most promise in constraining the cosmological model. These include: internal dwarf galaxy kinematics (rotation curves, stellar/gas/globular-clusters velocity dispersion); galaxy luminosity and stellar mass functions; satellite numbers and their luminosity functions;  radial distributions of satellites; morphology of dwarfs and Milky Way-like galaxies (frequency of bulges, thickness of disks); among others. In addition, such studies may be combined with other techniques which offer the possibility to constrain the population of ``dark" subhalos, or gravitationally collapsed halos that would not host any observable galaxy. Such efforts include studies of gravitational lensing \citep[e.g.,][]{1998MNRAS.295..587M,2002ApJ...572...25D,2009ApJ...699.1720K,2009arXiv0908.3001K,2010MNRAS.408.1969V}, gaps in stellar streams \citep[e.g.,][]{2011ApJ...731...58Y,2014ApJ...788..181N,2016ApJ...820...45C,2016PhRvL.116l1301B,2016MNRAS.463..102E} and limits from the Lyman-alpha forest \citep[e.g.,][]{2006PhRvL..97s1303S,2017PhRvL.119c1302I,2017PhRvD..96b3522I,PDB2020}. %\annika{tie more closely to ``observing" simulations.  Somewhere we do need to introduce observational systems that are promising, but need 3 might be the place that it first naturally occurs.} 
However, whether or not we can truly use these phenomena to constrain the physics that underpins our Universe depends critically upon whether or not we can actually observe them. Data taken from physically motivated, high resolution simulations and reduced using methods identical to those applied to observations are necessary to determine which observations are most likely to yield concrete answers. \\

\noindent \textbf{Summary:} Upcoming surveys linked to \textit{JWST}, the Vera C. Rubin Observatory, DES, and \textit{Roman}, among others, will dramatically increase our capability to detect and study galaxies -- in particular, the smallest dwarf galaxies in the universe, which are expected to be entirely DM-dominated. Theoretical models should, as much as possible, be prepared and upgraded in order to predict galaxy properties in an observation-friendly fashion, taking into account survey and telescope specifications. %\annika{good, I like this future focus.  Can we tie more closely to future demands for observing siulations?  I.e., what are the challenges and opportunities for these future data sets?  It would be helpful to delineate more strongly between current successes and challenges, and the challenges/opportunities of the future.} 
This means that the next decade will be a crucial period for collaboration between theorists and observers. Observers must share the pipelines being developed to reduce data from these next-generation instruments, as well as expected uncertainties, and theorists must be willing to examine not only the ground truth from their simulations, but also how that may differ (or cease to be visible at all) from an observer's point of view.

\subsection{Need \#5: Fast realization of observed systems for dark matter parameter constraints}\label{sec:need5}
%Writers: Daniel Gilman, Yao-Yuan Mao

\noindent \textbf{Current Status:} As discussed above, cosmological simulations with full hydrodynamics are a critical tool to reveal how different physical properties of DM alter the abundance and internal structure of DM halos and subhalos, which can result in observable differences in astronomical objects and systems. 
These simulations produces ``mock universes'' that allow us to compare theoretical prediction with observations in the space of observable. 
As such, running these simulations will become the bottleneck of parameter inference and model comparison, because these tasks typically require generating a large sample of simulated datasets of different input parameters (DM properties in this case).  

The speed with which simulated datasets can be generated in a forward model varies depending on the specific observational dataset used to test a particular theory. If the observational dataset requires a more sophisticated model to fully map out the space of observable, then more computational resource may be needed. 
For example, analyses of dwarf galaxies typically require a prescription to handle baryonic feedback inside Milky Way-sized halos \citep[e.g.][]{read2016,Tollet++16,Peirani++17}. Analyses of stellar streams, other the other hand, requires detailed  dynamical evolution of a stream inside a host halo with the presence of perturbations from dark subhalos \citep{Bovy++17,Banik++21}. Strong gravitational lensing can measure subhalo population of massive ellitptical galaxies, but simulating substructures in the $\sim 10^{13} M_{\odot}$ halos that tend to host early-type galaxies \citep{Moller:2003MNRAS.345....1M} can lead to significant increase in the computational expense \citep{Fiacconi++16}.

\noindent \textbf{Future Opportunities:} Here we identify a few methods that can lead to an increased efficiency of simulated structure formation processes for DM physics. 
Generally speaking, these methods can be classified into two categories: (1) reducing the computational cost of individual simulations by swapping some simulation components with models, and (2) reducing the number of simulations needed in our analyses. 

In the first category, the most commonly used approach is to reduce the computational expense of solving full hydrodynamics in cosmological simulations by using an empirical or semi-analytic model that are painted onto a lower-resolution, gravity-only simulation. This approach can significantly reduce the computational cost, often by 1--2 orders of magnitude or more. 
Semi-analytic models, some examples of which include {\tt{galacticus}} \citep{Benson++12}, {\tt{SatGen}} \citep{Jiang++20}, and {\tt{UniverseMachine}} \citep{Wang++21}\footnote{Additional examples include \cite{Zentner++05,Somerville++15}.}, bypass the computational expense of tracking individual particle orbits by identifying certain properties of halos, such their bound mass and tidal radius, and evolve them analytically. These models can either directly map halo properties into galaxy properties (or other observables), or generate high-resolution statistics (e.g., subhalo properties) based on low-resolution, large-scale environments \citep[e.g.][]{2019MNRAS.488.3143B,2020ApJ...893...48N,2105.12105,Kim:2021zzw}. While empirical or semi-analytic models can significantly increase the speed of running cosmological simulations, they require calibration against full hydro-dynamical simulations \citep{Pullen++14,Yang++20}. Once carefully calibrated, these models can rapidly generate representative samples of host halos with substructure, as is required for analyses of stellar streams and strong gravitational lenses. The predictions from semi-analytic models can be directly utilized by codes targeted to a particular aspect of structure formation targeted by a particular observational probe. 

In the second category, as discussed in \S\ref{sec:need1}, one way to reduce the number of simulations needed is by identifying a property of DM structure that does not depend on a particular DM theory, and interpret the results of experiments in terms of this quantity. This approach allows us to translate the constraint we obtain from a single measurement into constraints on a variety of DM theories.  For example, models of warm and interacting DM, as well as ultra-light fuzzy DM, predict a cutoff in the matter power spectrum on small scales, resulting in a paucity of of gravitationally bound halos below a characteristic mass \citep[e.g.][]{Schneider++12,Schive++16,Du++17,BoddyGluscevic18,nadler2021}. A single measurement of a minimum halo mass can therefore be interpreted in the context of warm and fuzzy DM, thus constraining two separate classes of DM theories with a single measurement. As another example, for models of self-interacting DM, the number of core collapsed halos as a function of halo mass can be interpreted in the context of a variety of self-interaction cross sections \citep{2021MNRAS.507.2432G}. 

Also in the second category, one can build emulators from a set of simulations with different input DM properties. The emulator will interpolate how the observables responds to the change in the input DM properties. With emulators, we eliminate the need to run new simulations during parameter inference. This emulation method has already been implemented to enable  constraints on, for example, the dark energy equation of state (EOS) and the total neutrino mass \citep[see e.g.][]{Heitmann2009,Heitmann2016,Wibking2019,2019ApJ...875...69D,2021JCAP...05..033P,2021ApJ...906...74T,2019JCAP...02..031R}. Recently, emulation techniques have also been applied to enable DM science \cite[see e.g.][]{2021PhRvL.126g1302R,2021PhRvD.103d3526R}. There is, however, a unique challenges in using emulators for DM physics: multiple emulators might be needed for different classes of DM, as they have very different input parameter spaces (e.g., WDM masses vs. SIDM cross sections), and also for different DM probes, as they have very different observable spaces (e.g., dwarf galaxies vs. strong lenses). 

Given the vast space of unconstrained DM theories, whose growth over the course of the last decade has outpaced the observational constraints, a combination of the aforementioned strategies will likely be required. One potential path forward proceeds as follows: First we generate a small suite of small-volume, high-resolution simulations with different DM properties and full hydrodynamics. Then we generate a larger set of larger-volume simulations, and apply empirical and semi-analytic models, calibrated against the first suite of full hydrodynamical simulations. Finally we build emulators based on the second suite of simulations.
These emulators will become the backbone of the analysis pipelines that are specific to the measurments (e.g., dwarf galaxies, streams, and strong lenses). In addition, these suites of simulations can also shed light on how best to set up experiments to target key observable quantities that distinguish $\Lambda$CDM from alternatives. \\

\noindent \textbf{Summary:} Generating cosmological simulations with full hydrodynamics and different DM properties will become the bottleneck of parameter inference and model comparison. Empirical or semi-analytic models can help reduce the computational cost of individual simulations.  Identifying macroscopic properties that can be used to constrain multiple DM models and building emulators can help reduce the number of simulations needed in our analyses. We will need to combine both approaches to cover vast space of unconstrained DM theories and the diversity of observational measurements.

% \begin{figure}
%     \centering
%     \includegraphics[clip,trim=0.5cm 4.5cm 0.2cm
% 		5cm,width=.95\textwidth]{DM_halo_populations_final.pdf}\vspace{10mm}
% 		\includegraphics[width=.95\textwidth]{Nadler2021PRL_Fig1.pdf}
%     \caption{\emph{Top panels}: Example figure for strong lensing that shows the multi-plane convergence in substructure, including subhalos and line-of-sight halos. The panels depict the same simulated population of halos in CDM (left), WDM with a mass function cutoff around $5 \times 10^7 M_{\odot}$ (center), and SIDM with a velocity-dependent cross section (right). The black line shows the critical curve for the lens system. \emph{Bottom panels}: Example figure, taken from Nadler et. al. 2021 \cite{nadler2021}. Projections of the subhalo population in a MW-like zoom-in simulation in CDM (left), in a WDM model applied to the CDM simulation (middle), and in a SIDM resimulation of this system (right). \ferah{Conceptual vverlap with Fig. 1, 3-- consult with others to create a joint figure?}}
%     \label{fig:lensinghalos}
% \end{figure}

% \begin{figure}
%     \centering
%     \includegraphics[width=.95\textwidth]{Nadler2021PRL_Fig1.pdf}
%     \caption{Example figure, taken from Nadler et. al. 2021. Projections of the subhalo population in a MW-like zoom-in simulation in CDM (left), in a WDM model applied to the CDM simulation (middle), and in a SIDM resimulation of this system (right). ...\ferah{overlap with Figs 1\&2-consult with others to create a joint figure?}}
%     \label{fig:satellite_distribution}
% \end{figure}

%\clearpage
\subsection{Need \#6: Provide guidance to observers about promising new signatures of dark matter physics}\label{sec:need6}
%Writers: Adrienne Erickcek, Benjamin Lehmann, Lina Necib

%DM particle theory has demonstrated numerous connections between DM microphysics and upcoming observables. 
\noindent \textbf{The Path to Future Opportunities via Current Status:} Given the complex relationship between DM properties at the microscopic level and cosmic observables, simulations can reveal unforeseen signatures of DM physics that generate new observational strategies. We illustrate this principle by describing recent simulation-based predictions that have proven important to interpret current observations and to prepare for upcoming ones. We also highlight opportunities for the next generation of simulations to establish novel observational probes of DM. %We divide future observables into four categories, roughly aligned with classes of simulations.

%\annika{For 2.6.1, I would frame this as something from sims that we were not expecting and that is really important for lab experiments.  Some of Besla's current work points in that direction to.  Surprises from sims that have implications for traditional lab experiments as well as astronomical observations.}

%EON suggested framing:  (ALE likes this wording and incorporated it into first paragraph)
%Given the complex relationship between dark matter properties and cosmic observables, simulations can reveal unforeseen dark matter signatures that feed back into observational strategies in a ``positive feedback loop.''  We illustrate this principle by describing recent simulation-based predictions that have proven important to interpret current observations and to prepare for upcoming ones. We also look ahead to synergies between the dark matter simulation and observational communities given the advances on both fronts expected in the next decade.

%\subsection{The Unique Dark Matter Distribution in the Milky Way}
%Combine current 2.6.1 with a paragraph highlighting the importance of the Large Magellanic Cloud for predicting and interpreting the Milky Way satellite population and the properties of certain stellar streams.  Include examples regarding the predicted impact on direct detection experiments.

% \subsubsection{Phase space distribution of DM in the Local Group}
\noindent \textbf{Example 1: The Unique DM Distribution in the Milky Way}

\begin{figure}
    \centering
    \includegraphics[width=.95\textwidth,trim={0 5.5cm 7cm 9cm},clip]{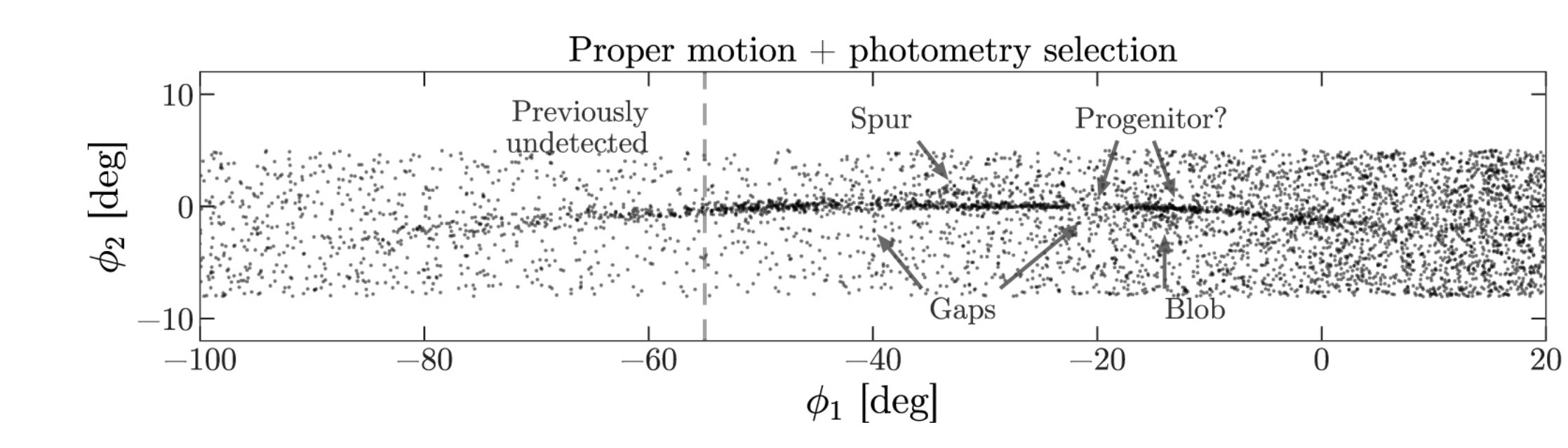}
    \caption{Gaps, spurs, and density inhomogeneities in the GD-1 stellar stream as observed by the {\it Gaia} satellite (taken from \cite{2018ApJ...863L..20P}). Such features are sensitive to DM substructure in the Milky Way.}
    \label{fig:stream_gaps}
\end{figure}

The DM phase space distribution encompasses two quantities: the DM density distribution and the DM velocity distribution.
Understanding
%the phase space distribution of DM
these distributions
within the Milky Way is crucial because (1) DM detection experiments are
%set on Earth, and therefore are
sensitive to the unique phase space distribution of DM in the solar neighborhood and throughout the Milky Way halo, and (2) observational deviations from $\Lambda$CDM, including the too-big-to-fail problem, the missing satellites, and the diversity problem often occur at Galactic scales (see \cite{2017ARAA..55..343B,Buckley:2017ijx} for reviews).
%Therefore, the study of DM distributions plays a central role in the future of DM detection. 

%Direct astrophysical probes provide a reasonably accurate measurement of the local density distribution (see \cite{2014JPhG...41f3101R} for a review). However, the local velocity distribution has not been directly measured. While it is often assumed to be a Maxwell--Boltzmann distribution \cite{Drukier:1986tm,Freese:1987wu},
%Studies using DM only simulations have shown that there are some small deviations from the Maxwell Boltzmann distribution \cite{2009MNRAS.395..797V,2010JCAP...02..030K,2012PhRvD..86f3505K}. this relies on the Milky Way having had a quiet merger history.
%With the observational advances of

Recent stellar catalogs, particularly \emph{Gaia} \cite{Gaia:2016zol,Gaia:2018ydn,2021A&A...649A...1G}, show that the Milky Way has had a more active and unique merger history than previously thought (see \cite{2020ARA&A..58..205H} for a review). Thus, it is essential to empirically measure the DM velocity distribution because standard assumptions of isotropy and equilibrium that lead to a Maxwell Boltzmann DM velocity distribution do not hold in detail. %\textbf{DM velocity distribution:} 
Refs.~\cite{2018PhRvL.120d1102H,2019ApJ...883...27N} introduced a new strategy for such a measurement, based on correlations between the stellar and DM velocity distributions. Such correlations were observed in the Eris and FIRE \textit{zoom-in hydrodynamic simulations of the Milky Way} \cite{2015MNRAS.450...53H,2018MNRAS.480..800H,2016ApJ...827L..23W} in the case of fully phase-mixed mergers.
%It is however crucial to further study such correlations using different baryonic prescriptions in hydrodynamic simulations.
However, these correlations only hold for CDM, and understanding the implications for different DM particle models requires further simulations implementing new microphysics on Galactic scales.

%\textbf{Stellar Streams:} 
Additionally, using stars as a proxy for the DM velocity distribution holds only in halos that include stars.  While massive satellites can contribute DM locally with no stellar tracer \cite{2019JCAP...11..013B}, it is important to separately study the contribution of dark subhalos (subhalos containing \emph{no} luminous baryons) using zoom-in simulations at different resolutions. In the Aquarius simulation \cite{2008MNRAS.391.1685S}, it was shown that dark subhalos might contribute a large fraction to the local DM density \cite{2011MNRAS.413.1373W}. Such dark subhalos can be probed for example via gaps in stellar streams.
Ref.~\cite{2018ApJ...863L..20P} has shown that the GD-1 stream \cite{2006ApJ...643L..17G} contains gaps (see Fig.\ \ref{fig:stream_gaps}) that could be caused by collisions with low-mass (and potentially baryon-free, or ``dark'') subhalos \cite{2019ApJ...880...38B,2020ApJ...892L..37B}.  %Ref.~\cite{2019ApJ...880...38B,2020ApJ...892L..37B} showed limits on the GD-1 perturber consistent
%within 3 standard deviations
%with $\Lambda$CDM subhalos.
%It is important to study such gaps in alternative DM models using simulations of stellar streams, as they could potentially be a novel source of constraints on particle DM.
DM microphysics can influence the population of such subhalos, and extracting predictions for the disruption of streams requires new simulations that implement non-minimal DM models on Galactic scales with substantial resolution.

Simulations that capture the unique assembly history of the Milky Way have also proven crucial to interpret observations of its substructure on larger scales. For example, the Milky Way's largest satellite galaxy, the Large Magellanic Cloud (LMC), is predicted to fall into the Milky Way with its own population of satellites, which has been characterized using zoom-in simulations and is required to explain the anisotropy in the observed satellite population after accounting for observational selection effects \cite{2020ApJ...893...48N,2021arXiv211204511M}. The LMC has also significantly perturbed several of the Milky Way's stellar streams, and idealized simulations including the dynamical impact of the LMC have been used to facilitate simultaneous inference of streams' orbital properties and the mass of the LMC \cite{2019MNRAS.487.2685E,2021ApJ...923..149S}. Simulations also predict that the presence of the LMC system may amplify certain signatures of DM physics at late times \cite{nadler2021}, induces global responses in the Milky Way halo that may be sensitive to DM properties \cite{2021ApJ...919..109G}, and impacts the interpretation of direct detection experiments \cite{2019JCAP...11..013B}. In parallel, observational efforts to map out satellite galaxies and stellar streams in the vicinity of the Magellanic Clouds (e.g., \cite{2021ApJS..256....2D}) and around LMC analogs (e.g., \cite{2016ApJ...828L...5C,davis2021,garling2021}) have recently intensified, creating new synergies between simulators and observers studying DM substructure on Galactic and sub-Galactic scales. These examples illustrate that the complexity of the DM distribution (particularly on small scales) in specific observed systems requires simulations to predict accurately, and that these predictions feed back into concrete observational strategies and synergies.

\noindent \textbf{Example 2: The Smallest DM Halos}

%\par\medskip\noindent
%\textbf{Minimum halo mass}: 
Many features of DM microphysics can influence the formation of small DM halos.
%Probing the population of such structures requires a detailed understanding of their baryonic tracers, which calls for a simulations implementing novel DM microphysics together with baryons at Galactic and sub-Galactic scales. 
%We highlight three observables of particular interest for DM microphysics given such a suite of simulations.
%Many non-minimal DM models produce modifications in the halo mass function.
%For example, famously, the free streaming of warm DM produces a cutoff at low halo masses.
%More broadly, a wide range of DM models with nontrivial effects on the matter power spectrum can be probed via reconstruction of the halo mass function. To enable this reconstruction
To infer the population of very low-mass halos from upcoming surveys, an extensive simulation program is needed to establish robust predictions for the underlying distribution of low-mass halos and the relationship between these systems and baryonic tracers. In addition to predicting the mass function, spatial distribution, and density profiles of the smallest halos in various DM models, simulations must determine the minimum halo mass for star formation at different epochs and characterize the impact of baryons on the extremely low-mass halo population.
%Basically current 2.6.3, with a broader framing that makes clear what simulation predictions for microhalos are novel/unexpected, and how that does/can/will inform experimental efforts to detect them.
%This could also incorporate a paragraph on the galaxy formation threshold and related insights from simulations, instead of the current Section 2.6.2. I'm not sure the Lyman-alpha forest recap is needed in 2.6.2. However, I worry that UDGs deserve a mention somewhere, and I'm not sure how they'd fit in with this proposal. Perhaps there could be  another subsection on "dark matter in anomalous systems" including UDGs, bullet cluster-like mergers, rare massive high-z galaxies, etc.
        
%\subsubsection{Microhalo detection}
\label{sims:mini}
DM halos well below the galaxy formation threshold are remnants of the first stages of structure formation that contain information about the early Universe and the properties of the DM particle. Enhancements to the primordial power spectrum \cite{Graham:2015rva} and early periods of matter domination (EMDE) \cite{Erickcek:2011us} or kination \cite{Redmond:2018xty} can trigger the formation of DM microhalos ($M \lesssim  0.01 M_\odot$) at redshifts greater than 200, long before they are expected to form under standard assumptions.  The size of the smallest halos may be determined by the free-streaming length of the DM particle, but it is also possible that other particles \cite{Blanco:2019eij, Erickcek:2020wzd,Erickcek:2021fsu} and the expansion history of the Universe \cite{Blanco:2019eij,Barenboim:2021swl} can influence the minimum halo mass.  In the case of axion DM, isocurvature fluctuations in the axion field generate axion miniclusters at high redshifts, with a minimum mass set by the horizon size when the Peccei-Quinn symmetry was broken \cite{Kolb:1994fi,2015PhRvD..92j3513G,Fairbairn:2017sil, Vaquero:2018tib,Buschmann:2019icd,Eggemeier:2019khm}. 

Sub-solar-mass microhalos are not expected to contain baryons; excitingly, simulations predict that these systems may nevertheless be observable.  If DM is self-annihilating, then microhalos boost the DM annihilation rate, leading to constraints on both the primordial power spectrum \cite{Bringmann:2011ut, Delos:2018ueo} and an early matter-dominated era (EMDE) \cite{Erickcek:2015jza,Erickcek:2015bda,Blanco:2019eij,StenDelos:2019xdk}.  Microhalos can also be detected gravitationally.  Upcoming pulsar timing arrays, like those that will be enabled by the Square-Kilometer Array (SKA; \cite{Weltman:2018zrl}), will be capable of detecting halos with masses as small as $10^{-13} M_\odot$ \cite{Dror:2019twh, Ramani:2020hdo, Lee:2020wfn, Delos:2021rqs} .  Sub-Earth-mass halos can also be detected using variations in light curves as stars pass near lensing caustics in galaxy clusters \cite{Dai:2019lud, Blinov:2021axd}.  

All of these detection strategies depend on the abundance and internal structure of the microhalos, and dedicated small-box microhalo simulations are required to generate robust predictions.  
%Recent simulations of axion minihalos \cite{Xiao:2021nkb} reveal that that the Press-Schechter mass function \cite{1974ApJ...187..425P} is not always accurate when applied outside of the context of standard CDM, and it does not describe how minihalos survive within larger halos.  
Simulations of microhalos subjected to tidal forces and impulsive energy injection indicate that early-forming microhalos are generally compact enough to survive tidal disruption \cite{Delos:2019lik} and stellar encounters \cite{Delos:2019tsl}, but 
%their survival depends on their density profiles.  The first halos that form in simulations with enhanced small-scale density perturbations have central cusps with $\rho \propto r^{-3/2}$ \cite{Gosenca:2017ybi,Delos:2017thv}, but this density cusp is softened to $\rho \propto r^{-1}$ by mergers between minihalos \cite{Ogiya:2016hyo, Delos:2019mxl}.   
simulations of microhalo interactions are needed to determine how mergers affect the microhalo population \cite{Delos:2019mxl}.  Furthermore, an EMDE can result in halo formation during the radiation-dominated era \cite{Blanco:2019eij}, and new simulations of microhalo formation are required to determine the abundance and density profiles of these halos.  New simulations are also needed to determine how the formation of halos during the EMDE inhibits the formation of microhalos after the EMDE \cite{Blanco:2019eij,Barenboim:2021swl}.  For axion DM, recent simulations have determined the minihalos' mass function and density profiles at $z\lesssim 19$ \cite{Eggemeier:2019khm, Xiao:2021nkb}, but simulations of axion minihalos within galactic tidal fields and simulations of stellar encounters are needed to determine how the minihalo population evolves after that time. All of these predictions will help determine observational strategies for detecting the signatures of the smallest halos, leading to exciting new connections between observers and theorists.

\noindent \textbf{Example 3: The Impact of DM Physics on Cosmological Observables}
%Broaden  the current Section 2.6.4 to include both points, and highlight examples of dark matter predictions that are surprising/unforeseen for either epoch; as one example, https://arxiv.org/abs/1708.04389.

%\subsubsection{Cosmological observables}
Large-volume simulations will help sharpen cosmological tests of DM physics that affect cosmological observables, allowing upcoming datasets to simultaneously probe DM physics and cosmology. One of the most pressing needs is to establish robust, simulation-based predictions for the connection between DM physics and the epochs of cosmic dawn and reionization. For example, energy released by annihilating or decaying DM can heat the intergalactic medium and contribute to reionization.  Upcoming probes of the Epoch of Reionization (EoR), such as HERA \cite{HERA:2021bsv} and SKA \cite{Weltman:2018zrl}, therefore have the potential to provide a new window into DM microphysics.  
%Given standard assumptions regarding structure formation, dark matter is constrained to play a relatively minor role in reionization, even when the annihilation boost due to halos is included \cite{Poulin:2015pna, Liu:2016cnk, Short:2019twc}, but light dark matter particles ($m_\chi \lesssim 100$ MeV) may leave a distinctive observable signature \cite{Lopez-Honorez:2016sur}.  Moreover, the early-forming minihalos discussed in Section \ref{sims:mini} would significantly enhance the dark matter annihilation rate during the EoR and could contribute to reionization long before the first astrophysical sources are expected to appear.

Thus far, most analyses of DM's potential role in the EoR have relied on semi-analytical calculations of the global energy injection from DM annihilation \cite[e.g.][]{Poulin:2015pna, Liu:2016cnk, Lopez-Honorez:2016sur, DAmico:2018sxd, Short:2019twc, Hiroshima:2021bxn}.  Hydrodynamical simulations are required to determine how DM halos introduce spatial variations in reionization; such simulations indicate that neglecting spatial variations in the DM annihilation rate underestimates the 21cm power spectrum and obscures a potential means of distinguishing between energy injection from DM and astrophysical sources \cite{List:2020rrj}.  However, these simulations assume that energy emitted by DM is immediately absorbed by the intergalactic medium, and that is not the case \cite{Slatyer:2015kla}.  Additional hydrodynamic simulations of reionization that include delayed deposition of the energy injected by DM and nonstandard structure formation scenarios are required to fully realize the potential of upcoming 21cm observations to probe DM.

Such large-volume simulations can also be used to refine probes of modified cosmologies. Cosmological discrepancies such as the $H_0$ tension have motivated numerous DM models that yield modified expansion histories, which in turn have implications for structure in the late universe \cite{Vattis:2019efj,Mau:2022sbf}. Among the most mature class of models features an early dark energy (EDE) component \cite{Poulin:2018cxd,Murgia:2020ryi}, and gross features of DM structure in these models have already been explored \cite{Klypin:2020tud}. At a minimum, establishing the best observational targets to test these models requires large scale DM-only simulations incorporating novel cosmologies. Such simulations will be most valuable in combination with studies of baryonic tracers in small-scale simulations, in addition to their inherent complementarity.\\

%subsubsection{Anomalous Systems as Probes of DM Physics}

%bullet cluster, ...

% \textbf{Ultra-diffuse galaxies:} UDGs are a promising target for comparison between simulations and observations. Recent observational catalogs have clarified the abundance of these objects and highlighted unusual features of their DM distributions \cite{XXX}, but the formation of these objects remains poorly understood. Simulation efforts have pointed to multiple formation channels with nontrivial effects in both DM and baryonic physics. Galactic-scale simulations incorporating baryons will be crucial if UDGs are to probe DM microphysics.

\noindent \textbf{Summary:} Simulations have revealed connections between the stellar and DM velocity distributions and the impact subhalos have on their galactic hosts, the ways DM microphysics affects the smallest baryonic structures, and how the DM halo population preserves a record of the universe’s evolution and the origins of DM. Further development of these novel probes of DM requires additional simulations at cosmological and galactic scales that incorporate alternatives to CDM, baryonic physics, and nonstandard cosmological histories, as well as small-box simulations of the formation and evolution of the first DM halos in multiple cosmological contexts.  Such specialized simulations will establish new synergies between astronomical observations and DM physics.%, they will also show us how to use DM to learn more about the universe.

% Cosmological parameter inference
%         \begin{itemize}
%             \item \textbf{Sim need:} big boxes
%             \item Early dark energy
%             \item Reionization / 21cm
%         \end{itemize}

% \textcolor{blue}{(I vote to drop the next paragraph -Lina)}
% \textbf{Diversity Problem:} One of the more recent problems on Galactic scales is that of the diversity of rotation curves \citep{2017PhRvL.119k1102K}. Some observations of the rotation curves of some dwarf galaxies have shown that galaxies of similar masses had a variety of rotation curves. One possible explanation for such behavior is self-interacting DM (see \cite{2018PhR...730....1T} for a review). In order to confirm this effect on structure formations, it is crucial to simulate a large set of dwarf galaxies with self-interacting DM, and seek to reproduce such behavior. A followup question would be whether this behavior can be also explained by other DM particle models. 

\section{Conclusion}

%\lipsum[3-4]

% Body of the text
Since the discovery of excess mass in clusters and galaxies decades ago \cite{Zwicky:1933gu,Rubin:1970zza,1978PhDT.......195B}, particle physics inspiration for, and constraints on, DM have been mediated from astronomical observations by cosmological simulations.  Forty years ago, when paired with observations, simulations showed that neutrinos could not be the dominant component of DM, and that something like cold DM described the universe on large scales \cite{White:1983fcs,Blumenthal:1984bp,Davis:1985rj}.  This observation fueled excitement for finding a DM candidate in newfangled supersymmetry theory \cite{Pagels:1981ke}, leading to years of work by particle physicists to characterize the physical effects of neutralinos and other particles in astronomical objects and laboratory searches \cite{Boveia:2018yeb,Schumann:2019eaa,Green:2021jrr,Slatyer:2021qgc}.  In the past twenty years, unexpected astronomical discoveries fueled a new wave of particle modeling building and inspired the next generation of ever-more-sophisticated simulations to connect the physics of DM and gas to observations \cite{Battaglieri:2017aum,Brooks2017}.  We use observational astronomy to measure \emph{where} the DM is.  Simulations tell us \emph{how} to translate this measurement to the particle properties of DM.  Using this combination of observation and simulation, particle physicists can tell \emph{what} DM is, and how to confirm its properties in the lab \cite{Battaglieri:2017aum}.  

A well-synthesized simulation, observational, theoretical, and experimental program is critical to revealing the nature of DM.  The challenge and opportunity for this decade is to develop a robust and vibrant simulations program that connects the ground-breaking capabilities of observational facilities \cite{DMRubinWP,DMDESIWP,DMCMB64WP,DMfacilitiesWP,TFastroprobesWP,Buckley:2017ijx} to an expanding ecosystem of particle models for DM and tailored lab experiments \cite{Battaglieri:2017aum}.  For a simulation program to be successful, simulators must work with particle physicists to identify promising particle DM models and map their phenomenology into a space of cosmological simulation parameters (Need \#1; \S\ref{sec:need1}).  The numerical algorithms to fold these new physics into simulation must be robustly tested (Need \#2; \S\ref{sec:need2}).  Importantly, simulations to test these models must include the physics of gas and stars---the past two decades of work in our community show that this is absolutely critical to making realistic predictions of observations as a function of new physics (Need \#3; \S\ref{sec:need3}).  As much as is possible simulations need to be ``observed" in the same way as the sky is---in the past, many discrepancies between theory and data arose because of differences in how theorists and observers were measuring properties of astronomical systems (Need \# 4; \S\ref{sec:need4}).  Even when simulations are well-tuned to interesting astronomical systems, there is a need to scale up their speed quickly in order to make DM parameter constraints using ensembles of observational and simulation data.  This is an important step to ``likelihood-function-izing" DM constraints from a wide set of astronomical observables, and connecting them to laboratory probes of DM (Need \#5; \S\ref{sec:need5}).  Finally, simulations play an essential role in developing new and promising pathways for astronomical constraints on DM (Need \#6; \S\ref{sec:need6}). 

The next decade will be game-changing in the DM community's ability to learn about DM in the sky and in the lab.  Simulations and the simulators who create them are the connectors between these two critical pathways to the discovery of the nature of DM.  A close collaboration among simulators, particle physicists, and observational astronomers is essential to the success of simulations to serve as the connector.  Only with a vibrant and cohesive cosmological simulation program will we be able to connect the lab to the sky to reveal the secrets of DM.

% References
\bibliographystyle{JHEP.bst}
\bibliography{main.bib}

\end{document}

%% file: intro_new.tex
Over the past century, various astronomical measurements, from a wide range of cosmological epochs and length-scales, have independently pointed to the existence of Dark Matter (DM). These measurements include galaxy rotation curves\footnote{see \cite{2019A&ARv..27....2S} for a recent review.} \cite{1978ApJ...225L.107R,1980ApJ...238..471R}, the Cosmic Microwave Background (CMB) (e.g., \cite{2011ApJS..192...18K,2020A&A...641A...6P}), and weak and strong gravitational lensing \cite{2006MNRAS.368..715M,2012ApJ...757...82B,2017MNRAS.468.1962N,2020MNRAS.491.6077G, 2020MNRAS.492.3047H}. Very little is currently known about the microphysical nature of DM, but its mere presence is an existence proof of physics beyond the Standard Model of particle physics. For decades, DM model building on the particle physics side focused on the Weakly Interacting Massive Particle (WIMP) paradigm. These models can be detected/constrained through a range of experimental techniques including collider, direct detection, and indirect detection experiments \cite{Boveia:2018yeb,Schumann:2019eaa,Slatyer:2021qgc}. To date, these experiments have been unable to detect the DM constituent, but have drastically reduced the allowed parameter space of WIMPs. 

In light of this, there is a major effort to move beyond the WIMP paradigm, driving research at the intersection of astrophysics, cosmology, and particle physics. Moving beyond the WIMP paradigm leads to some proposed particle physics models that can only be tested by astrophysical and cosmological probes rather than terrestrial experiments (see, e.g., \cite{Buckley:2017ijx}). They include the physics of DM or a dark sector that only communicates with the visible sector through gravitational effects; for instance, a warm DM candidate with non-negligible free streaming scale \cite{Dodelson:1993je,Baltz:2001rq}, fuzzy DM models \cite{2000PhRvL..85.1158H}, DM interactions with a dark thermal bath \cite{Feng:2009hw}, and DM self-interactions \cite{2000PhRvL..84.3760S}.

The microphysics of DM imprints itself onto the time evolution of the DM cosmic density in ways that are testable with a plethora of observations.  As summarized in the Snowmass contribution on halo mass \cite{DMhalosWP} and on astrophysical and cosmological probes of DM \cite{TFastroprobesWP}, different theories of DM affect the matter power spectrum and non-linear halo mass function in model-dependent ways, which can be tested at high redshift with the formation of the first stars, at moderate redshifts with gravitational lensing and galaxy luminosity functions, and at $z=0$ with ultrafaint satellite galaxies and stream gaps, among others.  Some models of DM affect the expansion history of the universe \cite{2002PhRvD..65f3506Z}, or do not change the expansion history but alter the growth function with respect to CDM, which may be testable using observations that span the cosmic microwave background to low-z probes of galaxies \cite{peter2010,2022arXiv220111740M}. Predicting the distribution of DM on sub-kpc scales matters for stellar probes of DM physics \cite{DMextremeWP,stellarsigWP} and for direct- and indirect-detection experiments \cite{2014PDU.....5...45P,StenDelos:2019xdk,Delos:2021rqs,2020PhRvD.101b3006O}. If DM consists at all of primordial black holes, then astronomical observations are the primary way we will learn about this class of model \cite{2021JPhG...48d3001G,DMPBHWP}.

These new ideas in DM coincide with the commissioning of various powerful observational facilities, including optical galaxy surveys, measurements of the CMB temperature, polarization, and lensing, and line-intensity mapping (see other Snowmass contributions on the subject, e.g., \cite{DMfacilitiesWP,DMRubinWP,DMCMB64WP,DMDESIWP}. This includes the Legacy Survey of Space and Time (LSST) by the Vera C.\ Rubin Observatory, planned to begin in 2023, and which will, among other things, provide the most complete census of the satellite population of the Milky Way galaxy \cite{drlica-wagner2019,DMhalosWP,DMRubinWP,TFastroprobesWP}. These new facilities promise sufficient raw sensitivity to provide powerful constraints on various classes of proposed DM models via many observational probes \cite{DMhalosWP,DMPBHWP,DMextremeWP,stellarsigWP,TFastroprobesWP}. To do so, however, requires a systematic method for mapping the microphysics of the particle physics models to the space of observables at these facilities. Since the difference in the phenomenology of these models is expected to be observable on scales and systems driven by nonlinear physics, robust modeling of these signals must necessarily be based on simulation-driven approaches, which can capture the full nonlinear physics. This simulation-driven approach is also crucial to understanding and disentangling the degeneracies between standard baryonic physics and new signals of DM physics \cite{Brooks2017,fitts2019}. It should also be stressed that such a simulation-driven approach will also enable other crucial aspects of the Cosmic Frontier --- including constraining the nature of Dark Energy, and constraining the total mass of the Standard Model neutrinos. These aspects are explored in greater detail in a companion Snowmass 2021 white paper \cite{compFrontierWP}.

To harness the full power of the new observational facilities in placing quantitative constraints on various classes of DM models, it is essential that particle physicists, simulation experts and observers collaborate to put a robust simulation and modeling program in place as the data starts coming in. In this white paper, we outline and detail the key needs of such a program, starting from appropriate initial conditions for the simulations, all the way to modeling the direct observables from the aforementioned facilities. For each need, we identify areas of growth needed both on the particle theory side and on the simulation implementation side, which should be prioritized moving forward. 

This white paper identifies the following priorities:\\
\begin{enumerate}
    
    \item {\bf Close collaboration between simulators and particle theorists} to both identify key models and areas of parameter space and to successfully implement these models. This includes starting from appropriate initial conditions to providing simulation outputs in a manner that is meaningful to particle theorists, simulators and observers alike.\\

    \item {\bf Algorithm development and code comparison tests} ensuring that simulations meet the required precision targets set by the sensitivity of the new facilities. For hydrodynamic modeling, this includes evaluation and comparison of different subgrid physics parameterizations. \\

    \item {\bf Performing simulations with full hydrodynamics with validated subgrid models and numerical resolution} at the relevant redshifts and cosmological scales.\\

    \item {\bf Analysis of outputs in the realm of observations} including mass functions, luminosity functions, galaxy morphology, kinematics, intracluster light all measured in an apples-to-apples manner with observations from current and upcoming facilities.\\

    \item {\bf Fast realizations of observables for inference of DM properties} in order to constrain DM particle parameters from observation on feasible timescales.\\

    \item {\bf Identifying novel signatures from simulations and guidance to observers} derived both from numerical simulations and fast realizations that point to signatures of DM physics. \\

\end{enumerate}

\begin{figure}
    \centering
    \includegraphics[width=.95\textwidth]{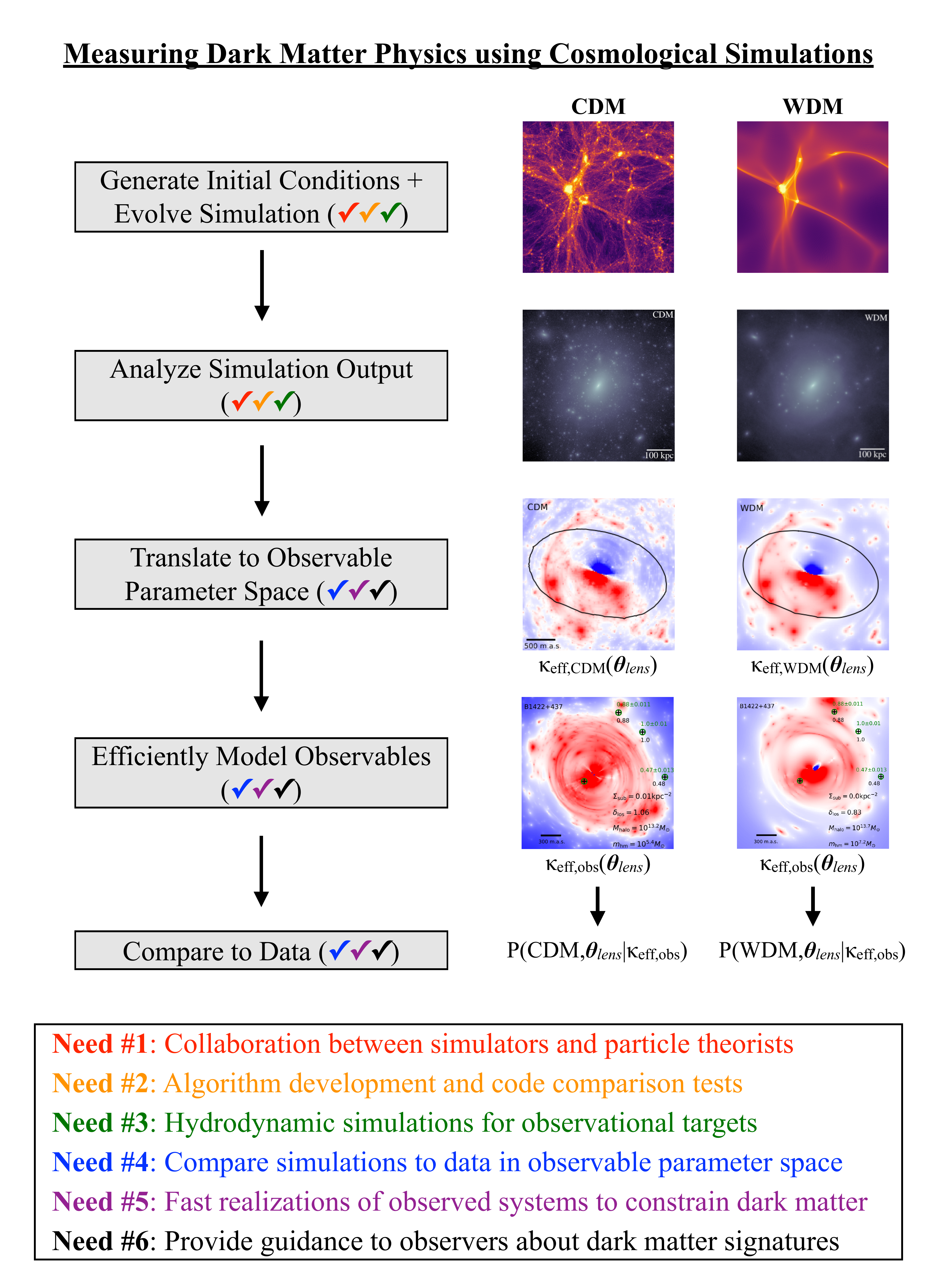}
    \caption{Flowchart for measuring DM physics using cosmological simulations and connections to the Needs outlined in this paper. Figures adapted from \cite{2017ARAA..55..343B,2020MNRAS.491.6077G} and provided by Philip Mocz, Daniel Gilman, Ethan Nadler.}
    \label{fig:flowchart}
\end{figure}

An example of a flowchart, showing how these six needs fit together for a single application in the context of substructure lensing, is shown in Fig.~\ref{fig:flowchart}.

%% file: need2.tex
%%   Writers: Ethan Nadler, Phil Mocz

% \pmocz{ Cover the classes of dark matter models that cosmological simulations cover (CDM, WDM, FDM, SIDM), and their numerical methods and initial condition generators \annika{consult with Need \#1} }

\begin{figure}
    \centering
{\footnotesize
\begin{tabular}{llll}
\hline
\mc{2}{\cellcolor{MyRed}\textbf{Cold Dark Matter (CDM)}}  & 
\mc{2}{\cellcolor{MyPurple}\textbf{Warm Dark Matter (WDM)}} \\
\cellcolor{MyRed} 
\begin{tabular}{l}
\includegraphics[width=3.4cm]{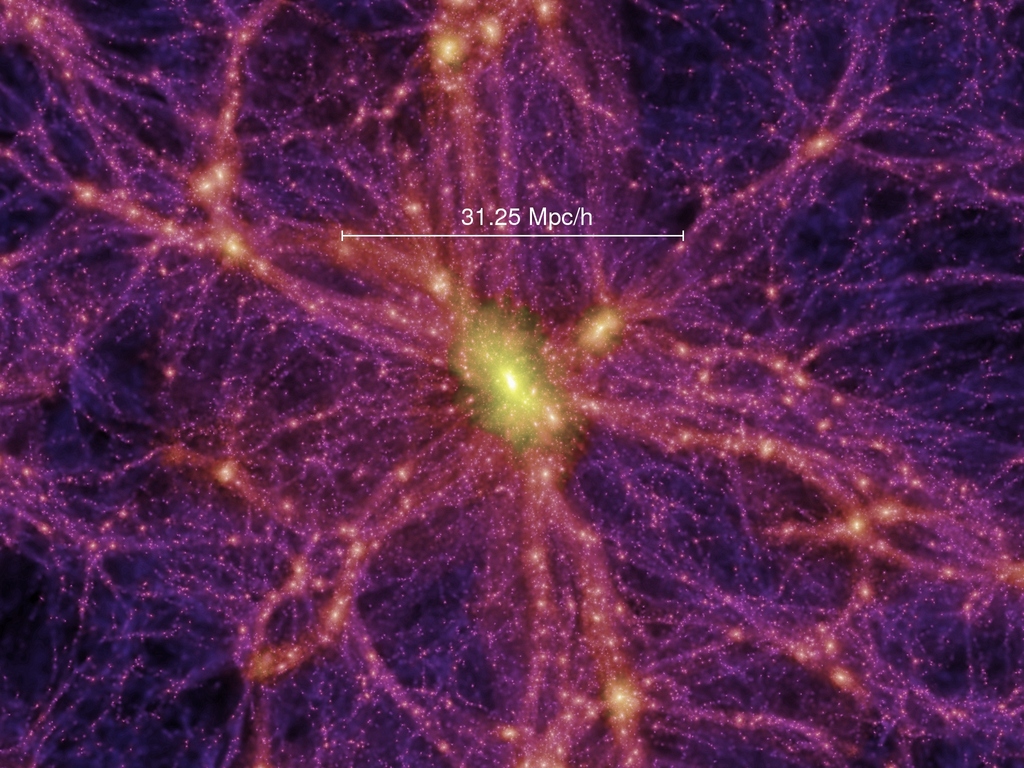} \\
Millennium simulation \\
\cite{2005Natur.435..629S}
\end{tabular} 
& \cellcolor{MyRed} 
\begin{tabular}{l}
WIMP \\
$\sim$100 GeV \\
collisionless \\
particle \\
\\
Vlasov-Poisson \\
equations
\end{tabular}
 &
\cellcolor{MyPurple} 
\begin{tabular}{l}
\includegraphics[width=3cm]{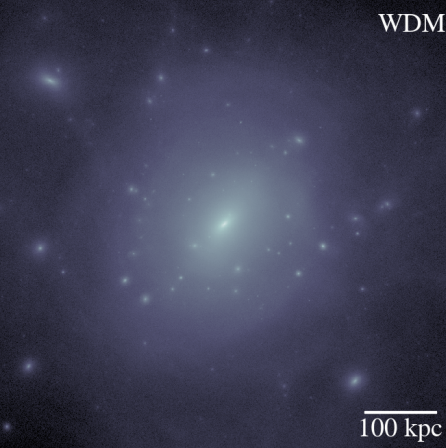} \\
WDM simulation \\
\cite{2017ARAA..55..343B} % 2017ARA&A..55..343B
\end{tabular} 
& \cellcolor{MyPurple} 
\begin{tabular}{l}
sterile neutrino \\
$\sim$3 keV \\
\\
Vlasov-Poisson \\
equations
\end{tabular}
\\
\cellcolor{MyRed} 
\begin{tabular}{|l|}
\mcl{1}{Numerical methods:}\\
\hline 
N-body: \\
$\cdot$ Tree, Fast Multipole \\
$\cdot$ {Particle-mesh/multigrid} \\
$\cdot$ Hybrid TreePM \\
Beyond N-body: \\
$\cdot$ Lagrangian tessellation \\
\hline
\end{tabular}
& \cellcolor{MyRed} 
 {\footnotesize
\begin{tabular}{|l|}
\mcl{1}{challenges:}\\
\hline 
softening-length \\ 
limits resolution, \\
Monte Carlo noise \\
\hline
\end{tabular} 
}
&
\cellcolor{MyPurple}
\begin{tabular}{|l|}
\mcl{1}{Numerical methods:}\\
\hline 
N-body methods: \\
with initial high-$k$ \\
exponential cutoff  \\
\hline
\end{tabular}
 & \cellcolor{MyPurple} 
 {\footnotesize
\begin{tabular}{|l|}
\mcl{1}{challenges:}\\
\hline 
artificial structure \\ 
from \\
`discreteness noise' \\
removed via\\
post-process \\
\hline
\end{tabular} 
}
\\
\mc{2}{\cellcolor{MyBlue}\textbf{Fuzzy Dark Matter (FDM)}} & 
\mc{2}{\cellcolor{MyYellow}\textbf{Self-Interacting Dark Matter (SIDM)}}  \\
\cellcolor{MyBlue} 
\begin{tabular}{l}
\includegraphics[width=3.4cm]{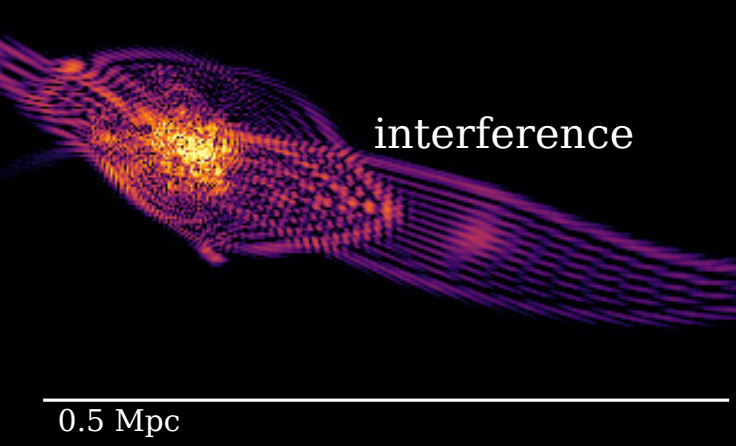} \\
FDM simulation \\
\cite{2019PhRvL.123n1301M}
\end{tabular} 
& \cellcolor{MyBlue} 
\begin{tabular}{l}
axion \\
$\sim10^{-22}$ eV \\
\\
Schr\"odinger-\\Poisson \\
equations
\end{tabular}
&
\cellcolor{MyYellow} 
\begin{tabular}{l}
\includegraphics[width=3.4cm]{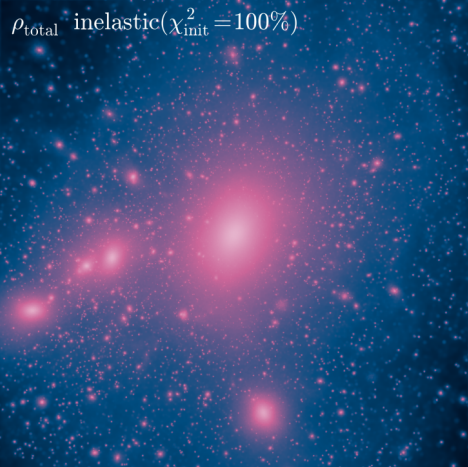} \\
SIDM simulation \\
\cite{2019MNRAS.484.5437V}
\end{tabular}
& \cellcolor{MyYellow} 
\begin{tabular}{l}
dark photon \\
$\sigma/m\sim {\rm cm/g}$ \\
collisional \\
particle \\
\\
Vlasov-Poisson \\
w/ collision
\end{tabular}
\\
\cellcolor{MyBlue} 
\begin{tabular}{|l|}
\mcl{1}{Numerical methods:}\\
\hline 
Spectral Methods \\
Finite Difference / AMR \\
Smoothed-Particle Hydro \\
\hline
\end{tabular}
& 
\cellcolor{MyBlue} 
 {\footnotesize
\begin{tabular}{|l|}
\mcl{1}{challenges:}\\
\hline 
Limited to small \\ 
cosmological \\
volumes \\
\hline
\end{tabular} 
}
& 
\cellcolor{MyYellow}
\begin{tabular}{|l|}
\mcl{1}{Numerical methods:}\\
\hline 
N-body methods: \\
with
Monte Carlo \\
probabilistic \\
scattering  \\
\hline
\end{tabular}
&
\cellcolor{MyYellow}
 {\footnotesize
\begin{tabular}{|l|}
\mcl{1}{challenges:}\\
\hline 
Limited scattering \\ 
cross-section types \\
resolved \\
\hline
\end{tabular} 
}
\\
\hline
\mc{4}{\cellcolor{MyGreen} \textbf{Multi-Physics}}  \\
\cellcolor{MyGreen}
{\footnotesize
\begin{tabular}{l}
Hydrodynamics \\
$\cdot$ Lagrangian SPH \\
$\cdot$ Eulerian AMR \\
$\cdot$ ALE moving mesh \\
$\cdot$ Mesh-free \\
\end{tabular} }
 &
\cellcolor{MyGreen} 
{\scriptsize
\begin{tabular}{l}
Gas Cooling \\
Interstellar Medium \\
Star Formation \\
Stellar Feedback \\
Supermassive-BHs \\
Active Galactic Nuclei \\
Magnetic Fields \\
Radiation Fields \\
Cosmic Rays 
\end{tabular} 
}
&
\cellcolor{MyGreen}
\begin{tabular}{l}
\includegraphics[width=3cm]{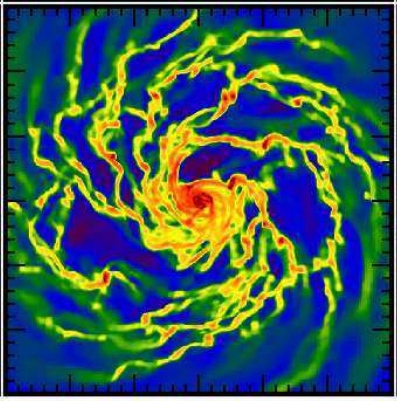} \\
AGORA simulation \\
\cite{2021ApJ...917...64R}
\end{tabular} 
&
\cellcolor{MyGreen} 
{\scriptsize 
\begin{tabular}{l}
\textbf{\footnotesize Initial Conditions} \\
 Primordial fluctuations \\
 initially evolved using \\
 perturbation theory \\
 \\
\textbf{\footnotesize Simulation Types} \\
Cosmological volume \\
Zoom-in
\end{tabular}
}
\end{tabular}
}
    \caption{ Cosmological simulation numerical methods and limitations for four types of DM models (CDM, WDM, FDM, SIDM), and common coupled multi-physics packages. }
    \label{fig:algorithms}
\end{figure}

%\annika{Some text overlap w/need 1--we can trim some of the first paragraph.  I like the focus from IC till today.} 
\noindent \textbf{Current Status:} %DM physics imprints its effects on structure formation over an extremely large dynamic range, including on highly nonlinear scales. 
As discussed in the previous section (\S\ref{sec:need1}), detailed cosmological simulations are necessary to accurately propagate the initial conditions and dynamical evolution for a given particle DM model to predictions for astrophysical and cosmological observables. Due to the computational expense of cosmological simulations, it is infeasible to simulate each DM scenario of interest; instead, the following phenomenological classes of DM models that capture its potential impact on structure formation are often considered\footnote{These classes are not comprehensive; for example, they do not include decaying DM, which is unstable with a lifetime comparable to the age of the universe and has previously been simulated \cite{2014MNRAS.445..614W,2021JCAP...10..040H,2022arXiv220111740M}, or DM--baryon interactions, which have been studied based on their suppression of the linear matter power spectrum \cite{2019ApJ...878L..32N,2021PhRvL.126i1101N,2021ApJ...907L..46M,2021PhRvD.104j3521N} and have only been simulated in specific cases \cite{2014MNRAS.445L..31B,2016MNRAS.461.2282S}.}, summarized additionally in Fig.~\ref{fig:algorithms}:

\begin{enumerate}
    \item \textbf{Cold DM (CDM)} is assumed to cluster down to extremely small scales and evolve as a collisionless fluid. CDM particles (e.g., WIMPs and QCD axions) decouple while non-relativistic and thus cluster down to mass scales as low as $\sim10^{-10}M_\odot$ \cite{Hogan:1988mp,Green:2005fa,2005PhRvD..71j3520L}. In most cosmological simulations, this extremely small free-streaming scale is not modeled or resolved (however, see e.g.\ \cite{2017MNRAS.471.4687A,2020Natur.585...39W}). Instead, the Vlasov-Poisson system that governs CDM is evolved in a cosmological background using N-body techniques, where each simulation ``particle" is assigned a mass many orders of magnitude larger than the DM particle mass and a gravitational potential softened on a corresponding length scale. Initial condition generators based on second-order Lagrangian perturbation theory \cite{2011MNRAS.415.2101H} and fast, massively parallelized N-body codes using tree-based \cite{2005MNRAS.364.1105S,2021MNRAS.506.2871S}, particle mesh, or hybrid \cite{2020ApJS..248...32W,2021MNRAS.508..575G} solvers are commonly used. Lagrangian techniques that track the evolution of the DM phase space sheet have also been developed in recent years \cite{2012MNRAS.427...61A}. Overcoming the impact of limited resolution to robustly resolve the smallest halos and subhalos remains a key challenge for CDM simulations \cite{2018MNRAS.475.4066V,2018MNRAS.474.3043V,2020MNRAS.491.4591E,2021MNRAS.505...18E,2021MNRAS.503.4075G}. %\annika{good to emphasize unique technical needs for each type of model.  Might be good to frame things in terms of current/past successes and future challenges.}
    
    \item \textbf{Warm DM (WDM)}  is assumed to feature a non-negligible free-streaming length but still evolve as a collisionless fluid \cite{1983ApJ...274..443B,2001ApJ...556...93B}. Initial conditions for WDM are usually calculated for a thermal relic particle that decouples while (semi)-relativistic, with a mass of $\mathcal{O}(1)$~keV. However, note that many well-motivated WDM models (e.g., sterile neutrinos) can feature non-thermal production \cite{2017JCAP...01..025A}. Structure formation in WDM is identical to that in CDM far above the free-streaming scale, and the N-body techniques described for CDM apply in these regimes. Resolution requirements are stringent for WDM simulations because spurious structures can artificially form if the numerical resolution and free-streaming scales are comparable \cite{2007MNRAS.380...93W}. Methods exist to identify and ``clean'' halo catalogs of spurious objects \cite{2014MNRAS.439..300L}, and phase spaced-based or hybrid methods can at least partially circumvent these issues \cite{2013MNRAS.434.3337A,2022MNRAS.509.1703S}. Detailed comparisons of the algorithms and simulation schemes used to mitigate spurious structure near the cutoff scale will be necessary to understand the systematic uncertainties associated with predictions from WDM-like simulations.
    
    \item \textbf{Fuzzy DM (FDM)} is assumed to consist of ultra-light $\mathcal{O}(10^{-22})$~eV particles and thus feature a non-negligible, galactic-scale  $\mathcal{O}(1)$~kpc de Broglie wavelength \cite{2000PhRvL..85.1158H,2017PhRvD..95d3541H}. Initial conditions for FDM are usually assumed to be dictated by interference effects alone---in particular, despite its extremely small particle mass, FDM is produced in a cold state, but it cannot cluster on small scales due to the uncertainty principle (this assumption is appropriate for many particle FDM models, e.g. ultra-light axions produced via the misalignment mechanism; \cite{2016PhR...643....1M}). FDM evolves according to the Schr\"odinger-Poisson equation, which imposes demanding spatial and temporal resolution criteria because both the complex amplitude and phase of the FDM field must be tracked \cite{2017MNRAS.471.4559M}. Several complementary methods for FDM simulations exist, including (pseudo)spectral solvers \cite{2017MNRAS.471.4559M,2018JCAP...10..027E,Du++17,2018PhRvD..97f3507D,2021PhRvD.104h3532G}, finite difference/adaptive mesh refinement schemes \cite{2014NatPh..10..496S,2016PhRvD..94d3513S}, and smoothed-particle hydrodynamics (SPH) approaches \cite{2015PhRvE..91e3304M,2018MNRAS.478.3935N,2021arXiv211009145S}. The Madelung transform of the Schr\"odinger-Poisson system is employed for the SPH approach to convert the wavefunction to a fluid equation, but this conversion faces challenges on resolving small scale interference substructure \cite{2015PhRvE..91e3304M}. All of these methods must be thoroughly compared and benchmarked on test problems to assess their accuracy.
    
    \item \textbf{Self-interacting DM (SIDM)} is assumed to interact with a non-negligible cross section of $\mathcal{O}(1)$~cm$^2$~g$^{-1}$ \cite{2000PhRvL..84.3760S,2018PhR...730....1T}. Although particle SIDM models can feature a suppression of the linear matter power spectrum (often with dark acoustic oscillations, as in e.g.\ dark photon mediator models) \cite{2016PhRvD..93l3527C}, many SIDM simulations only model the effects of self-interactions on the late-time evolution of DM. SIDM is typically assumed to interact elastically, with a velocity-independent cross section, though several simulations with inelastic and/or velocity-dependent interactions have been performed \cite{2012MNRAS.423.3740V,2013MNRAS.430...81R,2016MNRAS.461..710D,Huo:2019yhk,2019MNRAS.490.2117R,2020ApJ...896..112N,2021arXiv210608292B,nadler2021,2021MNRAS.500.1531C}. Self-interactions are simulated using modified N-body algorithms in which nearby simulation particles can exchange momentum with a probability determined by the interaction cross section and their relative velocity \cite{2012MNRAS.423.3740V,2017MNRAS.467.4719R,2020JCAP...02..024B}. These interactions between coarse-grained N-body particles operate over a much larger range than the corresponding microphysical interactions, and specific implementation choices may yield different predictions for evolution on small scales; thus, detailed code comparison studies will be necessary to ensure robustness \cite{2020ApJ...896..112N}. Recently, velocity-dependent SIDM models featuring core collapse in small halos have gained interest, both observationally \cite{2019MNRAS.490..231K,Sameie:2019zfo,2021MNRAS.503..920C,Yang:2021kdf,2021MNRAS.507.2432G} and theoretically \cite{2020PhRvD.101f3009N,2021MNRAS.505.5327T,2021arXiv211000259C}. Robust understanding and reliable numerical modeling of the phenomenon of core collapse in both idealized and cosmological simulations, will therefore, be necessary to enable precision predictions using SIDM simulations.

\end{enumerate}

%\pmocz{Gas-physics/hydrodynamics necessary to extract dark matter physics \annika{consult with Need \#3 and Need \#4}}
%\pmocz{Need to develop and validate algorithms for dark matter physics beyond the collisionless cold dark matter (CDM) \annika{consult with Need \#1}}
%\pmocz{Benchmarks for code comparison: (sub)halo mass function, density profiles, mass concentration relation, subhalo radial distribution, stellar-mass-halo-mass relation \annika{consult with Need \#5}}
%\pmocz{Comparison tools: halo finders, merger trees (need new algorithms for non-CDM?) \annika{consult with Need \#5}}

\noindent \textbf{Future Opportunities:} Simulations of the DM model classes described above must be validated to ensure that they reach the required precision for DM inference given the length scales and redshifts corresponding to cosmological observables of interest (see Needs 4-5). This validation should include the coupled (and potentially degenerate) effects of baryonic and DM physics, although DM-only simulations will continue to provide a useful benchmark given their interpretability and lower computational cost. Multiple kinds of DM phenomenology may appear simultaneously in specific particle models; for example, dark photon SIDM models also suppress the linear matter power spectrum and are effectively warm \cite{2018PhLB..783...76H}, and ultra-light axion FDM models can also feature significant self-interactions \cite{2015PhRvD..92j3513G}. Identifying theoretically motivated ``composite'' scenarios (Need 1; \S\ref{sec:need1}) to explore in simulations is an important area for development.

Key predictions to validate for each DM model class include:
The (sub)halo mass function, particularly on mass scales and in cosmological environments where precise small-scale structure measurements are feasible (e.g. within strong lenses and along their lines of sight, and within Milky Way analog halos); 
(Sub)halo density profiles, including their dependence on mass, environment, and DM properties, with descriptions of the mass–concentration relation as appropriate;
Subhalo radial distributions, particularly near the centers of host halos, where numerical biases can be severe even in CDM \cite{2021MNRAS.503.4075G}.
Additional subhalo population properties including infall time and phase-space distributions (e.g., velocity anisotropy profiles).
Currently, there are sizable systematics associated with simulations' predictions for these quantities, except (in some cases) for DM-only CDM simulations. Quantifying the level of precision the predictions must reach in each case, and performing code and algorithm comparison studies to ensure that these levels are reached, is therefore crucial.

For hydrodynamic simulations, different algorithms and sub-grid physics parameterizations can lead to significantly different predictions even for a fixed DM model (see Need 3; \S\ref{sec:need3}). Studying the interplay between DM microphysics and hydrodynamics, including star and galaxy formation physics, will be important for observables including ultra-faint dwarf galaxies and 21cm signals from cosmic dawn. Key predictions to validate include the stellar mass–halo mass relation and the galaxy occupation fraction. Baryons can also alter small-scale DM distributions (for example, through enhanced tidal stripping by central galaxies; \cite{2017MNRAS.471.1709G,2020MNRAS.492.5780R,2020MNRAS.499..116W,2022MNRAS.509.2624G}), and the efficiency of these processes may depend on the DM model class in question.

Finally, we emphasize that simulation analysis tools, including halo finders and merger tree algorithms, should be benchmarked for each DM scenario of interest. Code comparison studies have been performed in the CDM context that standard tools were developed in \cite{2011MNRAS.415.2293K,2012MNRAS.423.1200O,2013MNRAS.436..150S}, but have not been carried out in alternative DM scenarios and will be vital. This is particularly important because several models (e.g. WDM) are known to cause difficulties for standard halo finders \cite{2013MNRAS.434.3337A}, and spurious behavior may also arise for halos affected by self-interactions, fuzzy DM interference, or baryonic physics. %\annika{Good, I like this paragraph.  Think through what's the current barrier to doing these things, and how we might overcome them.}
\\

\noindent \textbf{Summary:} Detailed comparisons of simulation algorithms and analysis tools for benchmark cold, warm, fuzzy, and self-interacting DM, including in the presence of baryons, are necessary to respectively overcome challenges associated with limited numerical resolution, artificial fragmentation, robustly solving the Schr\"odinger-Poisson equation, and accurately resolving the effects of ram-pressure stripping and gravothermal collapse. Addressing these challenges through code comparison studies will enable theoretical predictions at the level of precision demanded by upcoming data.

% \pmocz{
% Have a figure summarizing dark matter models, methods, codes, types of simulations
% }

% Fig.~\ref{fig:algorithms}

%% file: need3.tex
% Writers: Stacy Kim, Mark Vogelsberger

\noindent \textbf{Current Status:} Many of the key discrepancies between predictions of $\Lambda$CDM and alternative DM models occur at small scales where baryonic physics plays an important --- if not fundamental --- role \citep{2017ARAA..55..343B}. At these scales, however, the impact of baryons is often degenerate with that of alternative DM models.  
For instance, the discrepancy between cuspy density profiles predicted by dark-matter-only simulations \citep{navarro1996} and observational hints of cores in some galaxies \citep[e.g.][]{walter2008, hunter2012}, can be explained both by popular alternative DM models such as SIDM \citep{2000PhRvL..84.3760S,Kaplinghat:2015aga,Ren:2018jpt} or FDM \cite{2000PhRvL..85.1158H}, as well as baryonic feedback from repeated starbursts \citep[e.g.][]{pontzen2012}, and other processes that induce potential fluctuations \citep[e.g.][]{el-zant2001, weinberg2007}.
Likewise, the number of observable satellite galaxies around galaxies like the Milky Way (MW) depends strongly on the suppression---if any---of low-mass subhalos, which can be induced by a suppression of power at small scales as in WDM%, the evaporation of subhalos due to late-time self-scatters as in SIDM, 
 as well as tidal stripping and destruction by the MW's baryonic disk.
These degeneracies pose a major challenge in producing robust DM constraints.

\noindent \textbf{Future Opportunities:} What is needed to break these degeneracies?  The complex interplay of the baryonic physics involved alone make simulations with full hydrodynamics essential.  Hydrodynamical simulations couple N-body methods to describe the DM component with hydrodynamical methods to model the gas dynamics. The latter methods can be divided into Lagrangian (e.g. smoothed particle hydrodynamics), Eulerian (e.g. adaptive mesh refinment) and arbitrary Lagrangian-Eulerian (e.g. moving mesh) methods.  In addition to the gas dynamics these simulations also employ different sub-resolution models for astrophysical processes that can not be directly resolved. This includes, among others: star formation, supermassive black holes, stellar feedback, active galactic nuclei, magnetic fields, cosmic rays, interstellar medium physics and cooling processes. The exact implementation of these sub-resolution models differs between different simulations and still remains a significant source of uncertainties \citep[e.g.][]{read2016, bose2019, munshi2019}. It is therefore crucial that different methods and implementations are compared against each other to understand systematic differences and how they impact the details of galaxy formation and its back reaction on the DM component. Further, in order to constrain these models, better observational tests of these sub-resolution models must be identified.  For example, gas kinematics \citep{el-badry2018} and stellar mass-metallicity relations in low-mass dwarf galaxies \citep{agertz2019} are sensitive to the strength of feedback, and may present a promising avenue to constrain feedback models.

While more accurate hydrodynamic CDM simulations will help constrain the extent to which baryonic physics can impact small scales, and simulations with alternative DM models, such as those highlighted in \S\ref{sec:need2}, will similarly constrain the extent DM physics can impact small scales, dedicated comparisons of fully hydrodynamic simulations \emph{with} alternative DM models will be key to breaking degeneracies between the two.  Studies in this vein have shown that while baryonic physics often cause galaxy properties in alternative DM models to converge with CDM, some differences remain.  For instance, while baryons can produce cores in SIDM, WDM, and CDM, only SIDM can produce cores in the smallest dwarfs \citep{robles2017, fitts2019}.  Further, different physics underlying core formation imprint different signatures onto stellar properties---SIDM galaxies are more extended \citep{vogelsberger2014}, have more isothermal velocity dispersions, and exhibit shallower stellar metallicity gradients than their CDM counterparts with cores induced by supernovae feedback \citep{burger2021}.  A similar study of subhalos of the MW % and the Large Magellanic Cloud 
indicate that baryonically-induced tidal stripping induces a mass-independent suppression of subhalo abundances, while SIDM and WDM induce a mass-dependent suppression, and that subhalo distributions are more isotropic in SIDM \citep{nadler2021}. Further simulations along this vein is needed. 

While we have highlighted the degeneracies between DM and baryonic physics for central densities and low mass halo abundances, the Vera C. Rubin Observatory, JWST, and the Nancy Grace Roman Space Telescope will usher in the discovery of many new types of systems with the potential to provide even sharper DM constraints.  These include the discovery of new stellar streams, strong lens systems, satellites of dwarf galaxies, isolated dwarf galaxies, and the first generation of stars and galaxies \citep{drlica-wagner2019}. %in sufficient numbers to enable precision tests of the difference between CDM and WDM mass functions at higher mass scales than the MW satellites
For many of these probes, detailed hydrodynamic simulations---much less detailed hydrodynamic simulations with alternative DM physics---are in their infancy (e.g. satellites of dwarf galaxies, the highly tidally stripped satellites of lens galaxies, or Milky Way streams with dwarf galaxy progenitors), or are non-existent (e.g. Milky Way streams with globular cluster progenitors and line-of-sight substructures).  Many of these promising DM probes share similar degeneracies between baryonic and alternative DM physics as with core formation and the abundances low-mass halos.  Hydrodynamic simulations will be key to establishing robust DM constraints for them as well. \\

\noindent \textbf{Summary:}  Upcoming surveys will be able to map out matter on unprecedentedly small scales.  These scales are heavily influenced not just by DM physics but also by baryons, which both produce effects that can mimic the other.  Hydrodynamic simulations will be key to exploring the interplay between the two and in identifying robust signatures of DM physics.